\documentclass[12pt,showpacs]{revtex4}
\usepackage[latin1]{inputenc}
\usepackage{graphicx}
\usepackage{amsmath}

\newenvironment{Glossarylist}{%
  \begin{list}{}{%
      \setlength{\labelsep}{1em}%
      }%
}%
{\end{list}}

\newcommand{\chem}[1]{\ensuremath{\mathrm{#1}}}
\newcommand{\un}[1]{\ensuremath{\unskip\,\mathrm{#1}}}

\begin{document}

\title[Capacitance fluctuations causing channel noise
reduction]{Capacitance fluctuations causing channel noise reduction
in stochastic Hodgkin-Huxley systems}
\author{G. Schmid\footnote[1]{Corresponding author:
Gerhard.Schmid@physik.uni-augsburg.de},
I. Goychuk and P. H\"anggi}

\affiliation{Institut f\"ur Physik, Universit\"at Augsburg,
D-86135 Augsburg, Germany}

\begin{abstract}

Voltage-dependent ion channels  determine  the electric properties
of axonal cell membranes. They not only allow  the passage of ions
through the cell membrane but also contribute to an additional
charging of the cell membrane resulting in the so-called capacitance
loading. The switching of the channel gates between an open and a
closed configuration is intrinsically  related to the movement of
gating charge within the cell membrane. At the beginning of an
action potential the transient gating current is opposite to the
direction of the current of sodium ions through the membrane.
Therefore, the excitability is expected to become reduced due to the
influence of a  gating current. Our stochastic Hodgkin-Huxley like
modeling takes into account both the channel noise -- i.e. the
fluctuations of the number of open ion channels -- and the
capacitance fluctuations that result from the dynamics of the gating
charge. We investigate the spiking dynamics of  membrane patches of
variable size and analyze the statistics of the spontaneous spiking.
As a main result, we find that the gating currents yield a drastic
reduction of the spontaneous spiking rate for sufficiently large ion
channel clusters. Consequently, this demonstrates a prominent mechanism
for channel noise reduction.

\end{abstract}

\maketitle
\section{Introduction}
\label{sec:introduction}

Following the study of Hodgkin and Huxley \cite{Hodgkin1952}, most
of the models of axons have treated the generation and propagation
of action potentials using deterministic  differential equations.
Since the work of Lecar and Nossal \cite{Lecar1971} it has become
increasingly evident, however, that not only the synaptic noise but
also the randomness of the ion channel gating itself may  cause
threshold fluctuations in neurons \cite{White2000,Koch}.
Therefore,
channel noise which originates in the stochastic nature of the ion
channel dynamics should be taken into account \cite{White2000,Koch}.
For example, in mammalian ganglion cells both the synaptic noise and
the channel noise might equally contribute to the neuronal spikes
variability \cite{Rossum}. Due to finite size, the origin of the
channel noise is basically due to fluctuations of the mean number of
open ion channels around the corresponding mean values. Therefore,
the strength of the channel noise is mainly determined by the number
of ion channels participating in the generation of action
potentials. Channel noise impacts, for example, such features as the
threshold to spiking and the spiking rate itself
\cite{Skaugen,Clay1983,Strassberg1993,DeFelice1993,Fox1994,Chow1996,Schneidman},
the anomalous noise-assisted enhancement of transduction of external
signals, i.e. the phenomenon of Stochastic Resonance
\cite{bezrukov_nature95, bezrukov_biophys97,Schmid2001,Jung2001,
Gammaitoni1998,Hanggi2002,Shuai03,Pustavoit03,schmidfnl,Gong05,
Jung06,Ya06} and the efficiency for synchronization
\cite{schmid2003}. Interestingly enough, there exist optimal patch
sizes for which the spike production becomes more regular, or the
response to external stimuli optimizes
\cite{Schmid2001,Jung2001,schmidfnl,Gong05,schmidphysbiol,Jung06}.
The objective of this work is to investigate the spontaneous spiking
of cell membrane patches when gating charge effects are also
considered.

When an ion channel opens or closes, an effective gating charge is moved
across the membrane \cite{Hille2001}. This motion creates
the so-called {\itshape gating current}
which is experimentally measurable
\cite{armstrongbezanilla73,keynesrojas73,armstrongbezanilla74}.
The influence of gating currents was not explicitly considered in the
original Hodgkin-Huxley (HH) model
\cite{Hodgkin1952}.
In 1975 Hodgkin \cite{hodgkin} and, independently, Adrian
\cite{adrian}, firstly inquired theoretically into the influence of
the ion channel density on the velocity of the action potential
propagation along the squid giant axon  by taking into consideration
the gating currents of sodium ion channels (via an effective
capacitance loading caused by ion channels, see also
\cite{Hudspeth}). Remarkably, they found an optimal ion channel
density for which the signal velocity is maximal
\cite{hodgkin,adrian,NossalLecarBook,adairpre,sangrey}. This led to
the emergence of the {\itshape Hodgkin's maximum velocity
hypothesis}, i.e. that the living species can adopt optimal
densities of ion channels to maximize the rate of neuronal
information transduction.

In this work we investigate the influence of both the gating
currents and the channel noise on the spontaneous spiking.
At the initiation of a spike, there occur two competitive currents
in the presence of ion channels: First, an ionic current, which is
caused by the influx of positively charged sodium ions through open
sodium ion channels into the cell, and second, a gating current that
flows in the opposite direction during the transition of sodium ion
channels from closed to open state.
The latter one is due to the movement of the positive gating charge
from the inside of the cell membrane to the outside. This current
actually precedes the first one, and terminates when the channel is
in its open state. However, due to spontaneous fluctuations between
the channel's states, a stochastic component of gating current also
emerges. Moreover, there are also gating currents due to potassium
channels which have been neglected within the macroscopic,
noise-free model \cite{hodgkin,adrian,NossalLecarBook,adairpre}.
These polarization currents of membrane-bound charges can
substantially contribute to the overall polarization current across
the membrane, which can be considered as an electrical capacitor.
Therefore, the
influence of gating currents on the excitability of small ion
channel clusters is {\it a priori} not clear and necessitates
further research.

In section~\ref{sec:shhmodel} we show how channel noise can be taken
into account within a Langevin equation approach. The next section
\ref{sec:gatingchargemodel} deals with the extensions to the
standard HH model with respect to gating charge effects. We
discuss the influence of gating charge on the dynamics with
respect to excitability due to a constant stimulus and as well as on
the deterministic spiking period. Thenceforth, in
section~\ref{sec:channelnoiseandgatingcharge} we consider the situation
where both channel noise and gating charge are considered and
demonstrate the influence on the mean interspike interval. The effect
of intrinsic coherence resonance in the presence of gating currents
is discussed in section~\ref{sec:coherenceresonance}.

\section{A stochastic generalization of the Hodgkin-Huxley model}
\label{sec:shhmodel}

According to the standard
HH model the dynamics of the membrane potential $V$
follows %
\begin{align}
   \label{eq:voltage-equation}
 C \frac{\mathrm{d}}{\mathrm{d}t} V +G_{\chem{K}}(n)\ (V-E_{\chem{K}})
   +G_{\chem{Na}}(m,h)\ ( V - E_{\chem{Na}})
    +G_{\chem{L}}
   \ (V - E_{L}) = I_{\mathrm{ext}}(t)\, ,
\end{align}
with the potassium and sodium conductances  given by:
\begin{equation}
  \label{eq:conductances-hodgkinhuxley}
  G_{\chem{K}}(n)=g_{\chem{K}}^{\mathrm{max}}\  n^{4} , \quad
  G_{\chem{Na}}(m,h)=g_{\chem{Na}}^{\mathrm{max}}\ m^{3} h\, .
\end{equation}
In Eq.~(\ref{eq:voltage-equation}), $V$ denotes the membrane
potential  measured throughout this work in [\un{mV}] and
$C=1\un{\mu F/cm^2}$ is the capacity of the cell membrane. The time
$t$ is scaled in \un{ms}. Furthermore,
$E_{\chem{Na}}=50\un{mV}$, $E_{\chem{K}}=-77\un{mV}$ and
$E_{\mathrm{L}}=-54.4\un{mV}$ are the reversal potentials for the
potassium, sodium and leakage currents, correspondingly. The leakage
conductance is assumed to be constant. $G_{\mathrm{L}}
=0.3\un{mS/cm^2}$, and {$g_{\chem K}^{\mathrm{max}}=36\un{mS/cm^2}$}
and {$g_{\chem{Na}}^{\mathrm{max}}=120\un{mS/cm^2}$} denote the
maximal potassium and sodium conductances, when all ion channels are
open. $I_{\mathrm{ext}}(t)$ indicates an external current stimulus.
Eq. (\ref{eq:voltage-equation}) is  nothing but a Kirchhoff law for
an electrical circuit made of membrane capacitor and variable
nonlinear conductances assuming that the conductances of {\it open}
ion channels are Ohmic, i.e. all the nonlinearity comes from their
gating behavior.

The gating variables $n,\ m$ and $h$, cf.
Eq.~(\ref{eq:voltage-equation}) and
Eq.~(\ref{eq:conductances-hodgkinhuxley}), describe the mean ratios
of the open gates of the specific ion channels. Assuming gate
independence, the factors $n^{4}$ and $m^{3}\ h$ are the mean
portions of the open ion channels within a membrane patch. The
dynamics of the gating variables are determined by voltage-dependent
opening and closing rates $\alpha_x(V)$ and $\beta_x(V)\; (x=m,h,n)$
taken at  $T=6.3 \un{^{\circ} C}$. They read
\cite{Hodgkin1952,goychukpnas,goychukphysicaa}: 
\begin{subequations}
\begin{align}
  \label{eq:rates-m}
  \alpha_{m}(V) &= 0.1 \, \frac{V+40}{1 - \exp\left\{-\, (V+40)/10
    \right\}}\, , \\
  \beta_{m}(V)  &= 4\, \exp\left\{ -\, (V+65)/18 \right\}\, , \\
  \alpha_{h}(V) &= 0.07  \, \exp\left\{ -\, (V+65)/20 \right\}\, , \\
  \beta_{h}(V)  &= \frac{1}{1 + \exp\left\{-\, (V+35)/10 \right\} }\, ,  \\
  \alpha_{n}(V) &= 0.01\, \frac{V + 55}{1 - \exp \left\{-\, (V + 55)
      / 10 \right\}}\, ,\\
  \label{eq:rates-n}
  \beta_{n}(V)  &= 0.125 \, \exp\left\{ -\, (V+65)/80 \right\}\, .
\end{align}
\end{subequations}
The dynamics of the mean fractions of open gates reduces in the standard HH
model to relaxation dynamics:
\begin{equation}
  \label{eq:gates}
  \frac{\mathrm{d}}{\mathrm{dt}} x =
  \alpha_{x}(V)\ (1-x)-\beta_{x}(V)\ x, \quad x=m,h,n\, .
\end{equation}
Such an approximation is valid for very large numbers of
ion channels and whenever fluctuations around their mean values are negligible.

However, each channel defines in fact a bistable stochastic element
which fluctuates
between its  closed and open states. %
The same is valid for the gates which assumed to be independent in the
HH model. The number of open gates undergoes a birth-and-death-like
process. The corresponding master equation can readily be written
down \cite{Fox1994}. The use of a Kramers-Moyal expansion in that
equation results in a corresponding Fokker-Planck equation which
provides a diffusional approximation to the discrete dynamics. The
corresponding Langevin equations then reads:
\begin{equation}
  \label{eq:stochasticgates}
  \frac{\mathrm{d}}{\mathrm{dt}} x =
  \alpha_{x}(V)\ (1-x)-\beta_{x}(V)\ x + \xi_x(t), \quad x=m,h,n\, .
\end{equation}
Here, $\xi_x(t)$ are independent Gaussian white noise sources of vanishing mean.
For an excitable membrane patch with $N_{\chem{Na}}$ sodium and $N_{\chem{K}}$
potassium ion channels the noise correlations assume the following form:
\begin{subequations}
\begin{eqnarray}
  \label{eq:correlator-a}
  \langle \xi_{m}(t) \xi_{m}(t') \rangle &= \frac{1}{N_{\chem{Na}}}\
   [\alpha_{m}(V)\, (1-m) + \beta_{m}(V)\, m]\ \delta(t -t')\, ,  \\
  \label{eq:correlator-b}
  \langle \xi_{h}(t) \xi_{h}(t') \rangle &=  \frac{1}{N_{\chem{Na}}}\
  [ \alpha_{h}(V)\, (1-h) + \beta_{h}(V)\, h]\ \delta(t -t')\, , \\
  \label{eq:correlator-c}
  \langle \xi_{n}(t) \xi_{n}(t') \rangle &=  \frac{1}{N_{\chem{K}}}\
  [ \alpha_{n}(V)\, (1-n)+
    \beta_{n}(V)\, n ]\ \delta(t -t')  \, .
\end{eqnarray}
\end{subequations}
The stochastic Eq. (\ref{eq:stochasticgates}) replaces Eq.
(\ref{eq:gates}). Note that the correlations of the stochastic
forces in these Langevin  equations contain the corresponding
state-dependent variables and thus should be $\rm It\hat
o$-interpreted because it is derived in a diffusional approximation of a
jump process which is intrinsically a white noise problem
\cite{Hanggi80, HanggiThomas,Zwanzig}, and does not result from a
white noise limit of a colored noise problem \cite{Hanggi95}.

Within the approximation of
homogeneous ion channel densities, {$\rho_{\chem{Na}} = 60\un{\mu
m^{-2}}$ and {$\rho_{\chem K} = 18\un{\mu m ^{-2}}$, the ion
channel numbers are given by: $N_{\chem{Na}}= \rho_{\chem{Na}} S,
\quad N_{\chem{K}}= \rho_{\chem{K}} S$, with $S$ being the size of
the membrane patch. The number of ion channels, or the
size of the excitable membrane patch $S$, respectively, determines
the strength of the fluctuations and thus the channel noise level.
With decreasing patch size, i.e.
decreasing number of ion channels, the noise level caused by
fluctuations of the number of open ion channels increases, cf.
Eq.~(\ref{eq:correlator-a})-(\ref{eq:correlator-c}). Within this
modelling the rate for spontaneous spiking depends monotonously on the
patch size \cite{Schmid2001,schmid2003,schmidfnl}.

\section{Gating charges and currents in a modified Hodgkin-Huxley model}
\label{sec:gatingchargemodel}

Our starting point for the derivation of the gating charge values are
the opening and closing rates,
Eq.~(\ref{eq:rates-m})-(\ref{eq:rates-n}), of the standard
HH model. Conformational changes of the ion channel in
order to switch between open and closed configurations come along
with movement of gating charge
\cite{Hodgkin1952,Hille2001}. Assuming Arrhenius-like
dependences, the
transition rates are \cite{Hille2001}:
\begin{subequations}
\begin{eqnarray}\label{alpha}
  \alpha (V) &= & \alpha_0 \exp \left\{ \frac{ q  z \cdot V}{k_{B}T} \right\}\, ,\\
  \label{beta}
  \beta (V) &=& \beta_0\exp \left\{ \frac{ - q  (1-z) \cdot V}{k_{B}T}
  \right\}\, ,
\end{eqnarray}
\end{subequations}
containing  the  voltage-independent parts $\alpha_0,\beta_0$, the
gating charge $q$, and the asymmetry parameter $z$ with $z \in
(0,1)$. The two latter parameters can be deduced by comparison of Eqs.
(\ref{alpha})-(\ref{beta}) with Eqs.
(\ref{eq:rates-m})-(\ref{eq:rates-n}) in the high activation barrier
limit, being assumed at large negative voltage values
\cite{goychukpnas,goychukphysicaa}, yielding \begin{subequations}
\begin{eqnarray}
  \alpha_{n}(V) &\propto \exp \left\{+\, V / 10 \right\}\, ,\quad  &
  \beta_{n}(V)  \propto \exp \left\{-\, V / 80 \right\}\, ,\\
  \alpha_{h}(V) &\propto \exp \left\{-\, V / 20 \right\}\, , \quad &
  \beta_{h}(V)  \propto \exp \left\{+\, V / 10 \right\}\, ,\\
  \alpha_{m}(V) &\propto \exp \left\{+\, V / 10 \right\}\, , \quad &
  \beta_{m}(V)  \propto \exp \left\{-\, V / 18 \right\}\, .
\end{eqnarray}
\end{subequations}
The corresponding gating charges then follow as: \begin{subequations}
\begin{eqnarray}
  q_{n} &= k_{B}T\, 1000 \cdot ( \frac{1}{10}+\frac{1}{80}) =
  2.709 e \, , \quad\\
  q_{h} &= k_{B}T\, 1000 \cdot ( -\frac{1}{20}+\frac{1}{10})=-3.612 e
  \, , \quad \\
  q_{m} &= k_{B}T\, 1000 \cdot ( \frac{1}{10}+\frac{1}{18})=3.746 e\,
  , \quad \label{qm}
\end{eqnarray}
\end{subequations}
with the elementary charge $e=1.6022\cdot 10^{-19} \un{A\cdot s}$.
In particular, $q_m$ and $q_n$ correspond to motion of effective positive charges during the gate openings
\cite{Hille2001} from the interior side of cell membrane to the exterior
one (outward motion).
The total gating charges are
\begin{subequations}
\label{eq:gatingcharge}
\begin{eqnarray}
  Q_{\chem{K}}^{\mathrm{n-charge}}  &= q_{n} \times 4\,  N_{\chem{K}}\, , \\
  Q_{\chem{Na}}^{\mathrm{m-charge}} &= q_{m} \times 3\, N_{\chem{Na}}\, , \\
  Q_{\chem{Na}}^{\mathrm{h-charge}} &= q_{h} \times 1\,  N_{\chem{Na}}\, .
\end{eqnarray}
\end{subequations}
To obtain the gating currents one has to multiply the gating charges
with time-derivatives of the corresponding gating variables. Adding
the corresponding current densities (per unit area) to the equation
for voltage variable (\ref{eq:voltage-equation}) we find
\begin{eqnarray}
  \label{eq:chargevoltageequation}
  I_{\mathrm{ext}}(t)= & C \frac{\mathrm{d} V(t)}{\mathrm{d} t} +
  g_{\chem{Na}}^{\mathrm{max}} \, m^{3}\, h \, ( V  - V_{\chem{Na}}) +
  g_{\chem{K}}^{\mathrm{max}} \, n^{4}\, ( V - V_{\chem{K}}) +
  g_{l}\, ( V - V_{l} ) \nonumber \\
  & + \rho_{\chem{Na}}\, 3\, \frac{\mathrm{d} m}{\mathrm{d} t}\,  q_{m}
  + \rho_{\chem{Na}}\,     \frac{\mathrm{d} h}{\mathrm{d} t}\,  q_{h}
  + \rho_{\chem{K}} \, 4\, \frac{\mathrm{d} n}{\mathrm{d} t}\,  q_{n}\, ,
\end{eqnarray}
where $\rho_{\chem{X}}$ are the specific ion channel densities. The
dynamics of the gating variables are given by
Eq.~(\ref{eq:stochasticgates}) and
Eq.~(\ref{eq:correlator-a})-(\ref{eq:correlator-c}). Taken together,
all these equations  constitute a stochastic generalization of the
HH model which accounts for gating current effects being
considered in the present work for the first time.

\begin{figure}[p]
  \centering
  \includegraphics{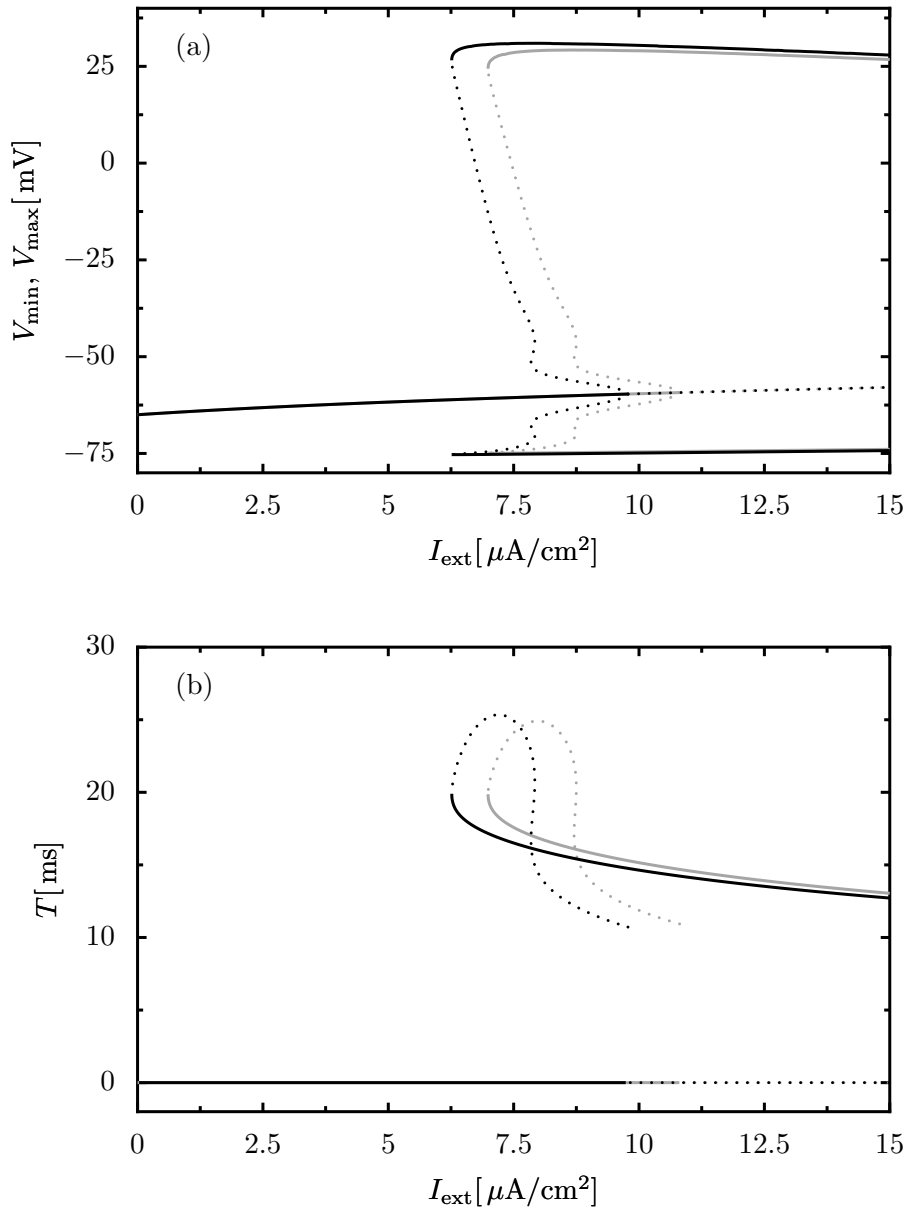}
  \caption{The bifurcation diagrams for the original deterministic HH
    model in (\ref{eq:voltage-equation}) (black lines) and the deterministic generalized model (grey
    lines) in (\ref{eq:chargevoltageequation}) which
    takes gating charge effects into account, are plotted versus
    the external constant driving current $I_{\mathrm{ext}}$.
    The continuous lines correspond to stable solutions, dotted lines
    refer to the unstable solutions.
    In Fig.~(a) the minimal and maximal membrane potentials are plotted.
    In Fig.~(b) the period of the spiking is depicted. A horizontal line
    at zero indicates the existence of a non-spiking solution.}
  \label{fig:bifurcation}
\end{figure}

\subsection*{Deterministic case: neglecting channel noise}

In order to study the influence of gating charge on the generation of
action potentials in more detail we consider the case of constant current driving,
i.e. $I_{\mathrm{ext}}(t)=const$. Fig.~\ref{fig:bifurcation} depicts
the bifurcation scenario for spike occurrence in the original HH
model and in the generalized model which takes gating charge into
account. As, for the moment, we want to neglect channel noise
(which corresponds to the limit of infinite patch size or infinite
numbers of ion channels),
the gating dynamics, Eq.~\ref{eq:stochasticgates}, reduces to that
in Eq. (\ref{eq:gates}).

While the existence of the fixed point solution is not affected by
the gating charge effects, both its stability range and the
emergence of the oscillatory spiking solution are shifted towards
larger values of the external driving current, cf.
Fig.~\ref{fig:bifurcation} (a). This again verifies the inhibitory
functionality of the gating charge. For the present case, the critical
current value at which the fixed point loses stability is
$I_{\mathrm{ext}}\approx 10.81 \un{\mu A/cm^{2}}$, whereas in the
standard HH model it occurs at $I_{\mathrm{ext}}\approx 9.76 \un{\mu
A/cm^{2}}$ \cite{hassard1978,rinzel1980}. Yet, before this happens
in both models in (\ref{eq:voltage-equation}) and in
(\ref{eq:chargevoltageequation}), a periodic, spiking solution
emerges via a subcritical Hopf bifurcation.  Then both the fixed
point solution and the limit cycle, i.e. the spiking solution, coexist
for a certain range of driving currents. The periods of the
solutions are depicted in Fig.~\ref{fig:bifurcation} (b). The
spiking period {\it vs.} the driving current  is also shifted
towards higher values of the driving current.

Fig.~\ref{fig:bifurcation}  demonstrates that the gating charge does
not affect the rest potential. It impedes, however, the system's
excitability. For a given $I_{\mathrm{ext}}$ the threshold for
excitation is increased, meaning that the excitability in the case
of a gating charge dynamics becomes decreased as compared to
 the case when no gating charge effects are present. In addition, the
time period of an oscillatory spiking solution is increased.

\subsection*{Deterministic case: increment of the membrane time constant }

At the beginning of an excitation the $m$ gates open, therefore
$\frac{\mathrm{d}
  m}{\mathrm{d} t}>0$. Consequently, the $m$-gating current is positive
(corresponding to an outward motion of positive charges), i.e.
$I_{\chem{Na}}^{\mathrm{m-gating}} = 3 \, N_{\chem{Na}}\, q_{m} \,
\frac{\mathrm{d} m}{\mathrm{d} t}>0$ ($q_{m}>0$).
The opening of
sodium channels enables transport of positively charged sodium ions
into the cytoplasm (implying a negative current by standard
convention).
Thus, these two currents (the gating or the polarization current
stemming from the ion channels and the ion current) are oppositely
directed. 
At the very beginning of the spike initiation (the
gating variable $m$ is ca. 0.1 at the threshold) the absolute value of
the gating  current is around 10 \% of the ionic
current.
The reason is that the gating current is proportional to the $m$
variable, whereas the averaged ion current scales with $m^3$.
Moreover, the gating current enhances the overall polarization
current across the membrane.
As a result, the membrane time constant
becomes effectively enhanced as it can be deduced from the following
reasoning:
To study the spike initiation, only the fastest gating variable
$m(t)$, which rapidly adjusts to the steady state value,
$m_{\infty}(V)=\alpha_m(V)/\left[\alpha_m(V)+ \beta_m(V)\right]$
should be taken into account. Note also that  $n(t)$ and $h(t)$
exhibit a slow dynamics and their values can be taken at the rest
potential $V_{\mathrm{rest}}=-65.0
 \un{mV}$. It then follows, using
 $\frac{\mathrm{d}m(V(t))}{\mathrm{d}t} \approx \frac{\mathrm{d}m_{\infty}(V)}{\mathrm{d}V}
\, \frac{\mathrm{d}V}{\mathrm{d}t}$, that the influence of gating
currents can be accounted for by a voltage-dependent contribution to
the membrane capacitance. Consequently, taking this $m$-gating
charge contribution into account, the effective membrane capacitance
reads $C_{\mathrm{eff}}(V)=C+C_{g}(V)$, where
\begin{align}
  C_{g}(V)= 3\, \rho_{\chem{Na}} \, q_{m} \,
  \frac{\mathrm{d}m_{\infty}(V)}{\mathrm{d}V} = 3\, \rho_{\chem{Na}}
  \, q_{m} m_{\infty}(V)\, \left[ 1 - m_{\infty}(V)\right]
  \,\frac{\mathrm{d}}{\mathrm{d}V} \, \ln \, \frac{\alpha_{m}(V)}{\beta_{m}(V)}
  \label{eq:capacitance}
\end{align}
This latter relation assumes a simple and insightful form in the
approximation of Eqs. (\ref{alpha}), (\ref{beta}):
\begin{eqnarray}
  C_{g}(V) \approx 3\rho_{\chem{Na}}\frac{q_m^2}{k_{B}T}
  m_{\infty}(V)[1-m_{\infty}(V)].
\end{eqnarray}
This expression evidences that $C_{g}(V)$ is maximal when the
sodium $m$-gates are half-open, i.e. $m_{\infty}=1/2$. For $q_m$
given in Eq. (\ref{qm}) and with $T=6.3 \un{^{\circ} C}$, we have
$C_{g} \approx 2.8\cdot 10^{-16}\, \rho_{\chem{Na}}
m_{\infty}(V)[1-m_{\infty}(V)]$ $\un{F}$. Furthermore, for
$\rho_{\chem{Na}}=60 \un{\mu m^{-2}}$, and for $m_{\infty}\approx 0.05$ at
the resting potential, $C_{g}(V_{\mathrm{rest}})\approx
0.085\un{\mu F/cm^{2}}$, i.e. the additional capacitance loading
caused by sodium channels amounts to about $8.5\%$ of the bare
membrane capacitance at the rest potential. However, it can
transiently be almost $42 \%$ of the bare capacitance when one-half
of the sodium gates is open. The membrane time constant at rest
$\tau_{\mathrm{rest}}=C_{\mathrm{rest}}/G_{\mathrm{rest}}$, where
$G_{\mathrm{rest}}$ is the corresponding membrane slope conductance,
is increased accordingly. Therefore, the spiking dynamics slows
down, cf. Fig.~\ref{fig:bifurcation}(b).

\section{Channel noise  and gating charge effects}
\label{sec:channelnoiseandgatingcharge}

The behavior predicted by the deterministic model should also carry
implications for the spontaneous spiking. Especially, we
expect a reduction of spiking activity due to gating charge effects for
membrane patches of finite size. This indeed is the case as we shall
demonstrate below.

One of the major consequences of intrinsic channel noise is
the initiation of spontaneous action potentials
\cite{White2000,Skaugen,Clay1983,Strassberg1993,DeFelice1993,Chow1996}.
A quantitative measure for the occurrence of the action potentials is
the mean interspike interval, i.e.,
\begin{equation}
  \label{eq:meaninterspike}
  \langle T \rangle = \frac{1}{N} \sum_{i=1}^{N} (t_{i}-t_{i-1})\, ,
\end{equation}
 where $t_{i}, \, i=1,..,N$ are the times for the
occurrence of spike events and we set  $t_{0}=0$. Since the strength
of the channel noise depends on the size of the membrane patch,
$\langle T \rangle$ is a function of the patch size $S$. With
increasing noise level or decreasing patch sizes $S$, respectively,
the spike production increases and thus the mean interspike interval
$\langle T\rangle$ decreases and even can approach the refractory
time \cite{Schmid2001,Jung2001}.

The spike occurrences $t_{i}$ are extracted from the voltage train
$V(t)$ which we obtain from numerical integration of the generalized HH model,
cf. Eq.~(\ref{eq:chargevoltageequation}), Eq.~(\ref{eq:stochasticgates})
and Eqs.~(\ref{eq:correlator-a})-(\ref{eq:correlator-c}). The
integration is carried out by a standard Euler algorithm with a step
size of $1\un{\mu
s}$. The Gaussian random numbers are generated by the ``Numerical
Recipes'' routine {\it ran2} using the Box-Muller algorithm
\cite{press1992}. To
ensure the confinement of the gating variables between $0$ (all
gates are closed) and $1$ (all gates are open) we have implemented
numerically reflecting boundaries at $0$ and $1$. The
occurrences of action potentials are determined by upward
crossings of the membrane potential $V$ of a certain detection
threshold. Due to the very steep increase of membrane potential at
firing, the actual choice of the detection threshold does not
affect the results. For each trajectory we performed $2^{28}$ simulation time
steps.
Moreover, an ensemble averaging over 100 different trajectories was done.

\begin{figure}[t]
  \centering
  \includegraphics{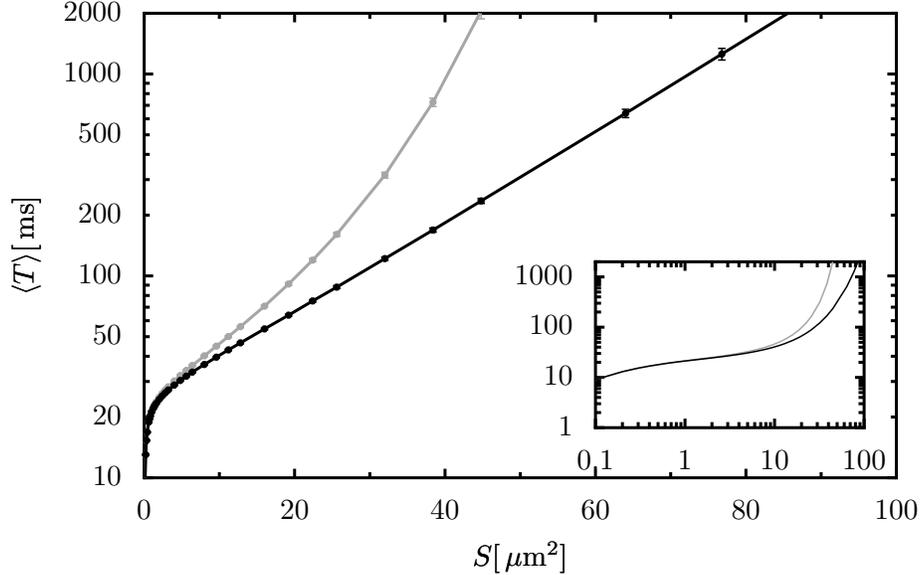}
  \caption{The dependence of the mean interspike interval $\langle T
    \rangle $ on the size $S$
    of the membrane patch  is depicted. The accuracy of the
    numerical simulation results is indicated by the error bars.
    If gating currents are
    considered, the spontaneous spiking activity, see the grey line,
    is reduced compared to the case for which gating charge effects
    are neglected, indicated by the black line. Especially, for large
    patch sizes or large numbers of ion channels, gating charge
    effects lead to a drastically reduced spontaneous spiking activity.
    In the inset, we depict the same data  on the log-log scale.}
  \label{fig:mean}
\end{figure}

In Fig.~\ref{fig:mean} we depict the dependence of the mean
interspike interval $\langle T \rangle$ on the patch size $S$. Due
to the inhibitory nature of the gating charge, the mean interspike
interval for a given membrane patch is significantly increased as
compared to the case where gating charge effects are neglected.
This leads to a
diminishing strength of the channel noise component, mostly
responsible for stochastic self-excitation. To put it differently,
stochastic fluctuations of the membrane capacitance partially
compensate the channel noise effect. Because the mean interspike
interval is exponentially sensitive to the channel noise intensity,
even a small relative reduction of noise intensity results in an
exponentially large effect for large membrane patches.
Therefore, the spontaneous spiking activity is drastically reduced.
For example, for patch sizes around $45\un{\mu m^{2}}$ the
probability of an occurrence of a spontaneous action potential is
one order of magnitude less in the case of gating charge effects as
compared to the case when gating charge effects are neglected. For
larger patch sizes the discrepancy becomes ever more striking. In
clear contrast, the gating charge effects can safely be neglected
for very small patch sizes for which the channel noise rules the
dynamics. Note that even though in this region a discrete stochastic
modeling is more appropriate, recent results show, that the
Langevin approach does work quite well also for small patch sizes
\cite{Jung06}.

\section{Coherence of Spontaneous Spiking Activity}
\label{sec:coherenceresonance}

Next we investigate the influence of the gating charge on the
regularity of the spontaneous spiking. A proper measurement is
provided by the coefficient of variation, $CV$, which is given as
the ratio of standard deviation to the mean value of the interspike
intervals, i.e.,

\begin{figure}[t]
  \centering
  \includegraphics{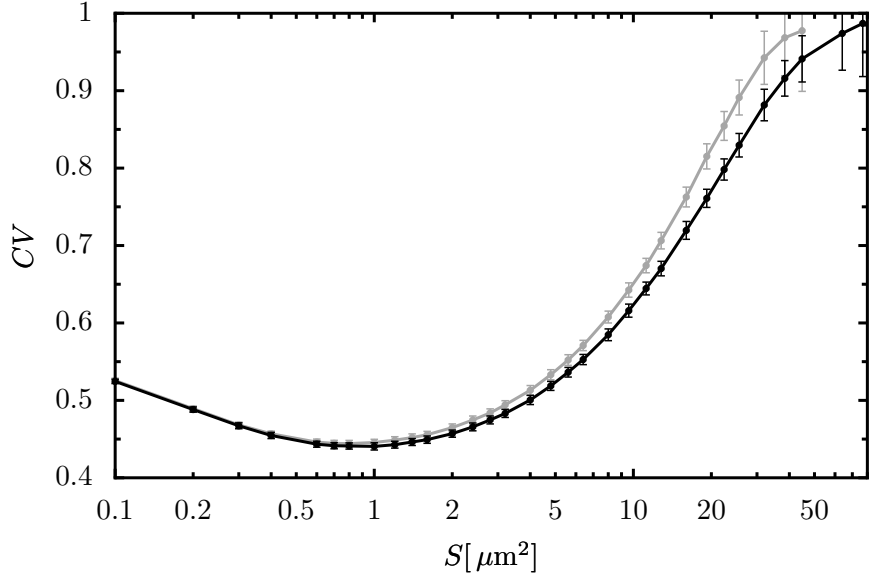}
  \caption{Same stochastic data as in Fig.~\ref{fig:mean}, but evaluated for  the coefficient of
    variation $CV$, cf. Eq.~(\ref{eq:cv}).
    }
  \label{fig:cv}
\end{figure}

\begin{equation}
  \label{eq:cv}
  CV = \frac{\sqrt{\langle T^{2}\rangle - \langle T
  \rangle^{2}}}{\langle T \rangle}\, ,
\end{equation}
with the mean squared interspike interval
 given by $\langle T^{2}\rangle
:= \frac{1}{N} \sum (t_{i}-t_{i-1})^{2}$. For a fully disordered
point process, meaning for a pure  Poisson process, the coefficient
of variation $CV$ assumes the value $CV=1$. For a more ordered
process $CV$ attains a smaller value, and for a deterministic signal
it vanishes.

Previous studies \cite{Schmid2001,Jung2001} where such gating
current effects have been neglected, have demonstrated that
the $CV$ exhibits a distinct minimum for an optimal patch size
$S\approx 1\un{\mu m^{2}}$. In our case again, the spiking is mostly
regular , cf. Fig.~\ref{fig:cv}. This phenomenon has been termed
intrinsic coherence resonance
\cite{Schmid2001,PikovskyKurths,Lindner04}. In Fig.~\ref{fig:cv},
the coefficient of variation is plotted {\itshape vs.} the patch
sizes for the two cases, with and without the gating charge effect
considered: the $CV$ shows a distinct minimum for both cases. Though
the most regular spiking could be observed at almost the same patch
size, namely $S=1\un{\mu}m^2$,  the spiking coherence slightly
deteriorates due to the gating currents' influence. The main effect
of gating charge on stochastic dynamics is thus to slow down the
spiking activity and this effect can be essential, cf. Fig.
\ref{fig:mean}.

\section{Conclusion and outlook}
\label{sec:Conclusion}

We extended a stochastic description of the  HH model accounting
for inherent channel noise by including gating current
effects, which equivalently can be described in terms of capacitance
fluctuations, see in eq. (\ref{eq:capacitance}). Our study revealed
that while the deterministic HH model with gating charge effects
does not differ dramatically from the original model for the
standard set of parameters, the corresponding stochastic model does
behave very differently for intermediate-to-large membrane patch
sizes. A main finding is that
 spontaneous spiking activity becomes drastically reduced. The
physical reason of this is that the gating current of translocated
gating charges in sodium channels is counter-directed to the
electrical current of sodium ions.
leading finally to excitation
after the whole system is driven by external current beyond a
threshold.
This results in the reduced channel noise intensity.
Simultaneously, there occurs a rise of the membrane time constant
resulting in longer spiking periods for supra-threshold driving.

For the parameters studied, i.e. for the original HH parameters, the
effects are relatively small in the noiseless case. Nevertheless,
even deterministically, gating charge effects lead ultimately to the
existence of  optimal ion channel densities for neuronal information
transduction \cite{hodgkin,adrian}.
Notably, an increase of  the sodium channel density by a factor of
$20$ (as in mammalian Ranvier nodes) would cause gating charge
effects of increased importance.
The
striking feature revealed in this work is that the gating charge
effects can play a fundamental role in the spontaneous spiking
dynamics of intermediate-to-large membrane patches even if they are
still relatively minor deterministically. The reduction of
spontaneous spiking exceeds one order of magnitude starting from a
patch size around $45 \un{\mu m}^2$ for the parameters studied. This
provides an inherent mechanism for the channel noise reduction in
neurons. This might explain the discrepancy between theoretical
predictions of stochastic HH model and some experimental results for
real neurons, e.g. see Ref.~\cite{Diba2004}.
However, the effect of intrinsic coherence resonance
as signature of the influence of channel noise in excitable
membranes still remains.

We share the confident belief that our investigations of the influence
of gating charge effects on the channel noise-induced spiking activity in
an archetypal model provide some new insights into the underlying physical
principles and mechanism of neuronal signaling.

\section*{Acknowledgments}

This work has been supported by the Deutsche
For\-schungs\-ge\-mein\-schaft via the Son\-der\-for\-schungs\-bereich SFB-486,
project A10.

\section*{Glossary}
\begin{Glossarylist}
  \item [Coherence Resonance] Noise-induced improvement of the
    regularity of the system's output. If the noise stems from
    internal dynamics the effect is called intrinsic coherence
    resonance.
  \item [Hopf-bifurcation]
    A change from a topology with fixed point solution to a topology
    with oscillatory solution under a small variation of a
     parameter within a nonlinear system.
  \item [It\^{o}-Stratonovich dilemma] Interpretation problem which arises
    in the context of Langevin equations in case of multiplicative
    Gaussian white noise.
  \item [Kramers-Moyal expansion] Procedure to transform
    a master equation for a discrete stochastic process into a Fokker-Planck
    equation for approximate
    continuous
    stochastic process.
  \item [Poisson process]  A stochastic process which is memory-less
    (Markovian) with exponentially distributed waiting times between
    two successive events.
   \item [Stochastic Resonance] A anomalous, noise-assisted enhancement
     of transduction of weak (deterministic or stochastic) signals.
\end{Glossarylist}

\end{document}